\documentclass[%
 reprint,
 amsmath,amssymb,
 aps,twocolumn
]{revtex4}

\usepackage{graphicx}
\usepackage{dcolumn}
\usepackage{bm}
\usepackage{subfigure}
\usepackage{color}
\usepackage{hyperref}

\begin{document}

\title{Maximally flexible  solutions of a random $K$-satisfiability formula}

\author {Han Zhao$^{1,2}$}
\email{zhaohan@itp.ac.cn}

\author {Hai-Jun Zhou$^{1,2}$}
\email{zhouhj@itp.ac.cn}

\affiliation{
$^1$CAS Key Laboratory for Theoretical Physics, Institute of Theoretical Physics, Chinese Academy of Sciences, Beijing 100190, China \\
$^2$School of Physical Sciences, University of Chinese Academy of Sciences, Beijing 100049, China
}

\date{\today}

\begin{abstract}
Random $K$-satisfiability ($K$-SAT) is a paradigmatic model system for studying phase transitions in constraint satisfaction problems and for developing empirical algorithms. The statistical properties of the random $K$-SAT solution space have been extensively investigated, but most earlier efforts focused on solutions that are typical. Here we consider maximally flexible solutions which satisfy all the constraints only using the minimum number of variables. Such atypical solutions have high internal entropy because they contain a maximum number of null variables which are completely free to choose their states. Each maximally flexible solution indicates a dense region of the solution space. We estimate the maximum fraction of null variables by the  replica-symmetric cavity method, and implement message-passing algorithms to construct maximally flexible solutions for single $K$-SAT instances.
\end{abstract}

\maketitle



\section{Introduction}
\label{intro}

The random $K$-satisfiability ($K$-SAT) problem is a paradigmatic model system of theoretical computer science~\cite{Mezard2009}. It has been widely adopted to understand the typical-case computational complexity of non-deterministic polynomial complete (NP-complete) optimization problems. It also serves as a convenient test ground for various empirical search algorithms. There are only two parameters: the number $K$ of variables involved in each constraint, and the ratio $\alpha$ between the number of constraints and the number of variables (the clause density). Phase transitions in this system has been extensively investigated in the statistical physics community following the initial empirical observations of Cheeseman, Kirkpatrick, and  colleagues~\cite{Cheeseman-Kanefsky-Taylor-1991,Kirkpatrick-Selman-1994} and the theoretical attempts of Monasson and Zecchina~\cite{Monasson1996,Zecchina1997}.

Deep insights have been achieved on the statistical properties of the random $K$-SAT solution space over the last two decades~\cite{Mezard2002,Mezard2002a,Montanari2004,Achlioptas-Naor-Peres-2005,Mezard2005,Mertens2006,krzakala2007,Montanari2008,Zhou2008,Achlioptas2008,Maneva2008,Zhou-Ma-2009,Zhou-Wang-2010,Lee-etal-2010}. It is now widely accepted that random $K$-SAT will experience a satisfiability phase transition as the clause density $\alpha$ exceeds certain critical value, $\alpha_{\rm s}$. The numerical value of $\alpha_{\rm s}$ as a function of $K$ can be  computed with high precision by the zero-temperature limit of the first-step replica-symmetry-breaking (1RSB) mean field theory of statistical physics~\cite{Mezard2002,Mezard2002a,Mertens2006}. Before the satisfiability phase transition occurs at $\alpha_{\rm s}$, the random $K$-SAT problem will first experiences several other interesting phase transitions as the clause density $\alpha$ increases, such as the emergence of solution communities~\cite{Zhou-Ma-2009,Zhou-Wang-2010,Li2009}, the breaking of ergodicity in the solution space (the clustering or dynamical transition)~\cite{Achlioptas-Naor-Peres-2005,Mezard2005,krzakala2007,Achlioptas2008,Maneva2008}, and the dominance of a sub-exponential number of solution clusters (the condensation transition)~\cite{krzakala2007,Montanari2008,Zhou2008}. Powerful message-passing algorithms, such as belief-propagation and survey-propagation, have been developed for solving hard random $K$-SAT problem instances~\cite{Mezard2002,Braunstein2005,Montanari-etal-ARXIV-2007,Marino-etal-2016}.

Most of the statistical physics studies in the literature consider solutions (satisfying configurations) that are picked uniformly at random from the whole solution space; in other words, every solution is assigned the same statistical weight and it has the same contribution to the partition function. This uniform statistical ensemble is appropriate for investigating typical configurations in the solution space, but it will completely miss all the different types of \emph{atypical} solutions. A recent surprising theoretical finding by Huang and co-authors was that the typical equilibrium solutions may be widely separated in the solution space as $\alpha$ becomes close to $\alpha_{\rm s}$, and then it should be very difficult to reach any of them by local dynamical processes~\cite{Huang-etal-2013,Huang-Kabashima-2014}. But empirical algorithms such as survey propagation did indeed succeed at $\alpha$ very close to $\alpha_{\rm s}$~\cite{Mezard2002,Braunstein2005}. Some recent reports suggested that atypical solutions are very important for understanding the performance of empirical $K$-SAT algorithms~\cite{Dallasta2008,Li2009,Zeng-Zhou-2013,Krzakala-Mezard-Zdeborova-2014,Baldassi2015,Baldassi2016,Budzynski2019}. Especially, Baldassi and co-authors found that there are sub-dominant and dense clusters in the $K$-SAT solution space, and biasing the search process towards such clusters can greatly increase the chance of solving a given $K$-SAT problem instance~\cite{Baldassi2015,Baldassi2016}. 

We follow this research line on atypical solutions in the present work, and discuss the issue of maximally flexible solutions. A  maximally flexible solution has the property that it satisfies all the constraints of a $K$-SAT instance only using the minimum number of variables. An extensive number $N \rho_0$ of the $N$ variables in such a configuration can be deleted while the formula is still satisfied. These insignificant variables are referred to as the null variables and $\rho_0$ is the fraction of null variables. Such atypical solutions have high internal entropy (at least of order $2^{N \rho_0}$), because the null variables are completely free to choose their states. Therefore, each maximally flexible solution is associated with a dense region of the solution space. We estimate the maximum fraction of null variables by the  replica-symmetric (RS) cavity method~\cite{Mezard2001}, and our theoretical results suggest that the maximum fraction $\rho_0^{\rm max}$ of null variables is positive at the satisfiability phase transition point $\alpha_{\rm s}$. We implement two message-passing algorithms to construct maximally flexible solutions for single $K$-SAT instances. More work needs to be done to extend the theoretical and algorithmic investigations to the 1RSB level.

The paper is organized as follows: In Sec.~\ref{sec:theory} we introduce a three-state model and write down the RS mean field equations. We then describe in Sec.~\ref{section:BPD} and Sec.~\ref{section:BPRf} two message-passing algorithms and discuss the numerical results. Finally we conclude this work in Sec.~\ref{section:concl}.

\section{Replica-symmetric mean field theory}
\label{sec:theory}

\begin{figure} 
\centering
\includegraphics[width=0.8\linewidth]{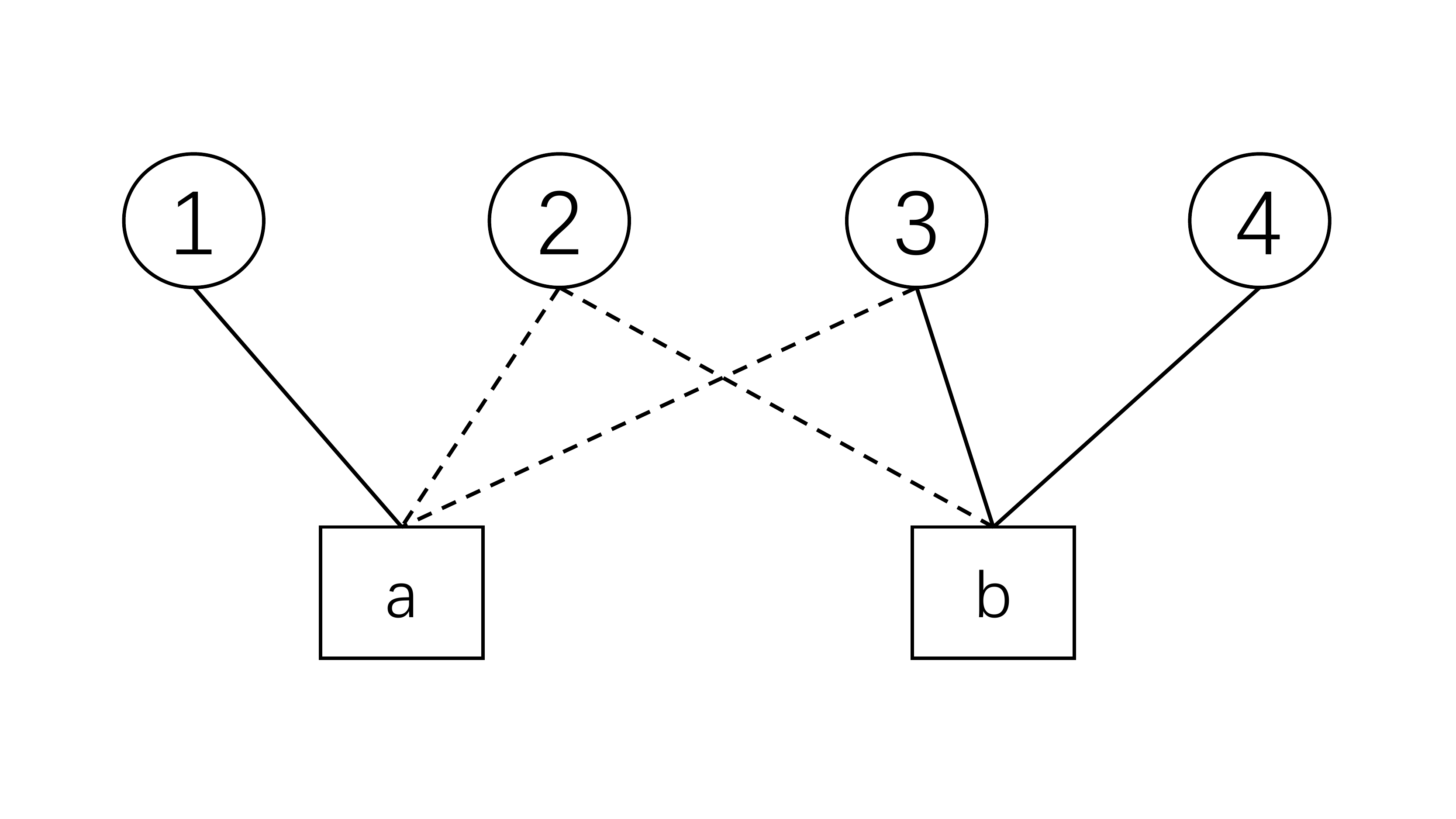}
\caption{Factor graph representation for a $K$-SAT formula composed of $N = 4$ variables and $M = 2$ clauses. The variables and clauses are denoted by circular and square nodes, respectively. Each clause has $K = 3$ attached links in this example. Each variable $i$ has three states ($s_i \in \{-1, 0, +1\}$) in our generalized model. The energy $E_a$ of a clause $a$ is either zero or unity. A solid link $(i, a)$ between variable $i$ and clause $a$ means a positive coupling constant ($J_a^i =1$) and that $E_a = 0$ if $s_i = 1$, while a dashed link means a negative coupling constant ($J_a^i = -1$) and that $E_a = 0$ if $s_i = -1$. If variable $i$ takes the null state $s_i = 0$, it does not contribute to satisfying any of the attached clauses.}
\label{fig:1}
\end{figure}

There are $N$ variables in a $K$-SAT formula (instance) and these variables are subject to $M$ constraints (clauses). The clause density $\alpha$ is defined as
\begin{equation}
    \alpha \equiv \frac{M}{N} \; .
\end{equation}
A simple $K$-SAT formula instance with $K=3$ is shown in Fig.~\ref{fig:1}, which means
$$
(x_1 \vee \overline{x_2} \vee \overline{x_3}) \wedge (\overline{x_2} \vee x_3 \vee x_4) \; ,
$$
where $x_i \in \{{\tt TRUE}, {\tt FALSE}\}$ is the Boolean state of variable $i$, $\overline{x_i}$ denotes the Boolean negation of $x_i$; and $\vee$ and $\wedge$ denote the Boolean {\tt OR} and {\tt AND} operators. There are $N = 4$ variables and $M \
= 2$ clauses, so the clause density is $\alpha = 2$.

In the original $K$-SAT problem each variable $i$ can only take two states $\sigma_i = -1$ (corresponding to Boolean {\tt FALSE}) and $\sigma_i = +1$ (corresponding to Boolean {\tt TRUE}), similar to the Ising model, so the total number of possible microscopic configurations is $2^N$~\cite{Monasson1996}. Each clause $a$ involves $K$ variables (we consider $K=3$ and $K=4$ in this paper), and its energy $E_a$ is either zero (clause satisfied) or unity (clause unsatisfied):
\begin{equation}
\label{eq:Ea}
    E_a = \prod\limits_{i \in \partial a} \frac{1 - J_a^i \sigma_i}{2} \; ,
\end{equation}
where $J_a^i \in \{-1, +1\}$ is the quenched binary coupling constant between clause $a$ and variable $i$ ($J_a^i=-1$ if variable $i$ is negated in clause $a$, otherwise $J_a^{i}=+1$), and $\partial a$ denotes the set of variables participating in clause $a$ (the cardinality $| \partial a | = K$). The clause energy $E_a$ will be zero if any of the variables $i \in \partial a$ takes the state $\sigma_i = J_a^i$. A factor-graph representation for the $K$-SAT problem is shown in Fig.~\ref{fig:1}, where each clause is linked to $K$ variables.

In the random $K$-SAT problem the $K$ variables of each clause $a$ are chosen uniformly at random from the $N$ variables, and the link coupling constants $J_a^i$ are independently assigned the value $-1$ or $+1$ with equal probabilities~\cite{Mezard2002}. The average number of attached links to a variable is $K \alpha$.

In this work we extend the random $K$-SAT problem by allowing each variable $i$ the possibility of a null state. The state of variable $i$ in the modified system is denoted as $s_i$ and it can be $+1$, $-1$ or $0$ (the null state). If $s_i = 0$ then variable $i$ does not contribute to satisfying any of the attached clauses, and these clauses need to be satisfied by its other connected variables. The modified energy $\tilde{E}_a$ of clause $a$ is then
\begin{equation}
    \tilde{E}_a =  \prod\limits_{i \in \partial a} \bigl( 1 - \delta_{s_i}^{J_a^i} \bigr) \; ,
\end{equation}
where $\delta_m^n$ is the Kronecker symbol, $\delta_m^n = 1$ if $m=n$ and $\delta_m^n = 0$ if $m \neq n$.

A microscopic configuration of the extended $K$-SAT system is denoted as $\underline{s}\equiv (s_1,s_2,...,s_N)$. There are a total number $3^N$ of possible configurations, but in this work we only allow those which satisfy the given $K$-SAT formula, that is, the global constraint is the zero-total-energy condition
\begin{equation}
   \sum\limits_{a=1}^M \tilde{E}_a = 0 \; . 
\end{equation}
The partition function for the system is then
\begin{equation}
\label{eq:Zbeta}
Z(\beta) = \sum_{\underline s} \prod\limits_{i=1}^{N} e^{\beta \delta_{s_i}^0}
\prod\limits_{a=1}^{M} \Bigl[1 - \prod\limits_{j \in \partial a} \bigl( 1-\delta_{s_j}^{J_a^j} \bigr) \Bigr] \; .
\end{equation}
Here a positive inverse temperature $\beta$ is introduced to encourage more null variables.

Following the replica-symmetric (RS) cavity method of statistical mechanics, which assumes that the variables participating in a given clause $a$ will become mutually independent if this clause is deleted from this system (i.e., the Bethe-Peierls approximation for a locally tree-like factor graph~\cite{Mezard2009}),  we write down the belief-propagation (BP) equation for the partition function (\ref{eq:Zbeta}) as
\begin{eqnarray}
q_{i\to a}(s_i) & = & \frac{e^{\beta\delta_{s_i}^0} \prod\limits_{b\in\partial i\backslash a} p_{b\to i}(s_i)}
{\sum\limits_{s_i^\prime} e^{\beta\delta_{s_i^\prime}^0} \prod\limits_{ b \in\partial  i\backslash a} p_{b\to i}(s_i^\prime) } \; ,
\label{a} 
\\
p_{a \to i}(s_i) & = & \frac{1-(1-\delta_{s_i}^{J_a^i})\prod\limits_{j\in\partial a\backslash i}[1-q_{j\to a}(J_a^j)]}
{3-2\prod \limits_{j\in\partial a\backslash i}[1-q_{j\to a}(J_a^j)]} \; .
\label{b}
\end{eqnarray}
Here $q_{i\to a}(s_i)$ is the cavity probability that variable $i$ would take state $s_i$ in the absence of clause $a$; $p_{a\to i}(s_i)$ is the cavity probability that variable $i$ would take state $s_i$ if it only participates in clause $a$; the set $\partial i$ contains all the clauses to which variable $i$ are linked, and $\partial i \backslash a$ means excluding clause $a$ from this set $\partial i$ (and similarly, $\partial a\backslash i$ means excluding variable $i$ from the variable set $\partial a$).

The marginal probability $q_i(s_i)$ of variable $i$ being in state $s_i$ is then evaluated as 
\begin{equation}
q_{i}(s_i) = \frac{e^{\beta\delta_{s_i}^0} \prod\limits_{{a\in\partial i}} p_{a\to i}(s_i)}{\sum\limits_{s_i^\prime} e^{\beta\delta_{s_i^\prime}^0} \prod\limits_{a\in\partial i} p_{a\to i}(s_i^\prime)} \; .
\label{q}
\end{equation}
The mean fraction $\rho_0$ of null variables at a given value of inverse temperature $\beta$ is computed through
\begin{equation}
\rho_0 = \frac{1}{N} \sum\limits_{i=1}^N q_i(0) \; .
\end{equation}
We are aiming at constructing $K$-SAT configurations which contain a maximum number of null variables. To estimate the maximum fraction $\rho_0^{\rm max}$ of null variables achievable for a given problem instance, we will compute $\rho_0$ as a function of inverse temperature $\beta$ using the above expression.

To determine the maximal value of $\beta$ that is physically meaningful, we also need to compute the free energy and entropy of the system. The total free energy $F(\beta)$ is defined as
\begin{equation}
 F(\beta) \equiv - \frac{1}{\beta} \ln Z(\beta) \; .
 \end{equation}
 In the RS mean field theory $F(\beta)$ can be decomposed into two parts, the contributions $f_{i+\partial i}$ of the variables $i$ and the associated clauses, and the contributions $f_a$ of single clauses $a$~\cite{Mezard2001,Mezard2009}:
\begin{equation}
F(\beta)  = \sum\limits_{i=1}^N f_{i+\partial i} - \sum\limits_{a = 1}^M (K - 1) f_a \; .
\end{equation}
The minus sign before the second summation in the above expression can be understood as follows: each clause $a$ is considered $K$ times by the contributions $f_{i+\partial i}$ of its $K$ attached variables $i$, so there should be a correction term $(K-1) f_a$. The expressions $f_{i+\partial i}$ and $f_a$ are, respectively, 
\begin{eqnarray}
 & & \hspace*{-0.5cm} f_{i+\partial i}  =  -\frac{1}{\beta} \ln \Big\{ e^\beta \prod\limits_{a\in \partial i} \Bigl[ 1 - \prod\limits_{j\in \partial a\backslash i} [1 - q_{j\to a}(J_a^j)] \Bigr] + \nonumber \\
& & \sum\limits_{\sigma_i= \pm 1} \prod\limits_{a \in \partial i} \Bigl[1 - \frac{1-\sigma_i J_a^i}{2}
\prod\limits_{j\in \partial a\backslash i} [1 - q_{j\rightarrow a}(J_a^j)] \Bigr] \Bigr\} \; , 
 \\
& & \hspace*{-0.5cm} f_a  =  -\frac{1}{\beta}\ln\Big[1-\prod_{j\in\partial a}[1-q_{j\to a}(J_a^j)]\Big] \; .
\end{eqnarray}
The free energy density is then $f \equiv F(\beta)/N$, and the entropy density $s$ of the system is computed through
\begin{equation}
s = - ( \rho_0 + f ) \beta \; .
\end{equation}
The entropy density $s$ measure the abundance of satisfying configurations with null variable fraction $\rho_0$. The value of $s$ may saturate to a positive value as $\beta$ increases (and $\rho_0$ also saturates to the limiting value $\rho_0^{\rm max}$), or it may become negative as $\beta$ exceeds certain threshold value $\beta_{\rm max}$. In the latter case we take the computed value of $\rho_0$ at $\beta_{\rm max}$ as the maximum null variable fraction $\rho_0^{\rm max}$.

\begin{figure}
\centering
\subfigure[]{
\includegraphics[width=3.0in]{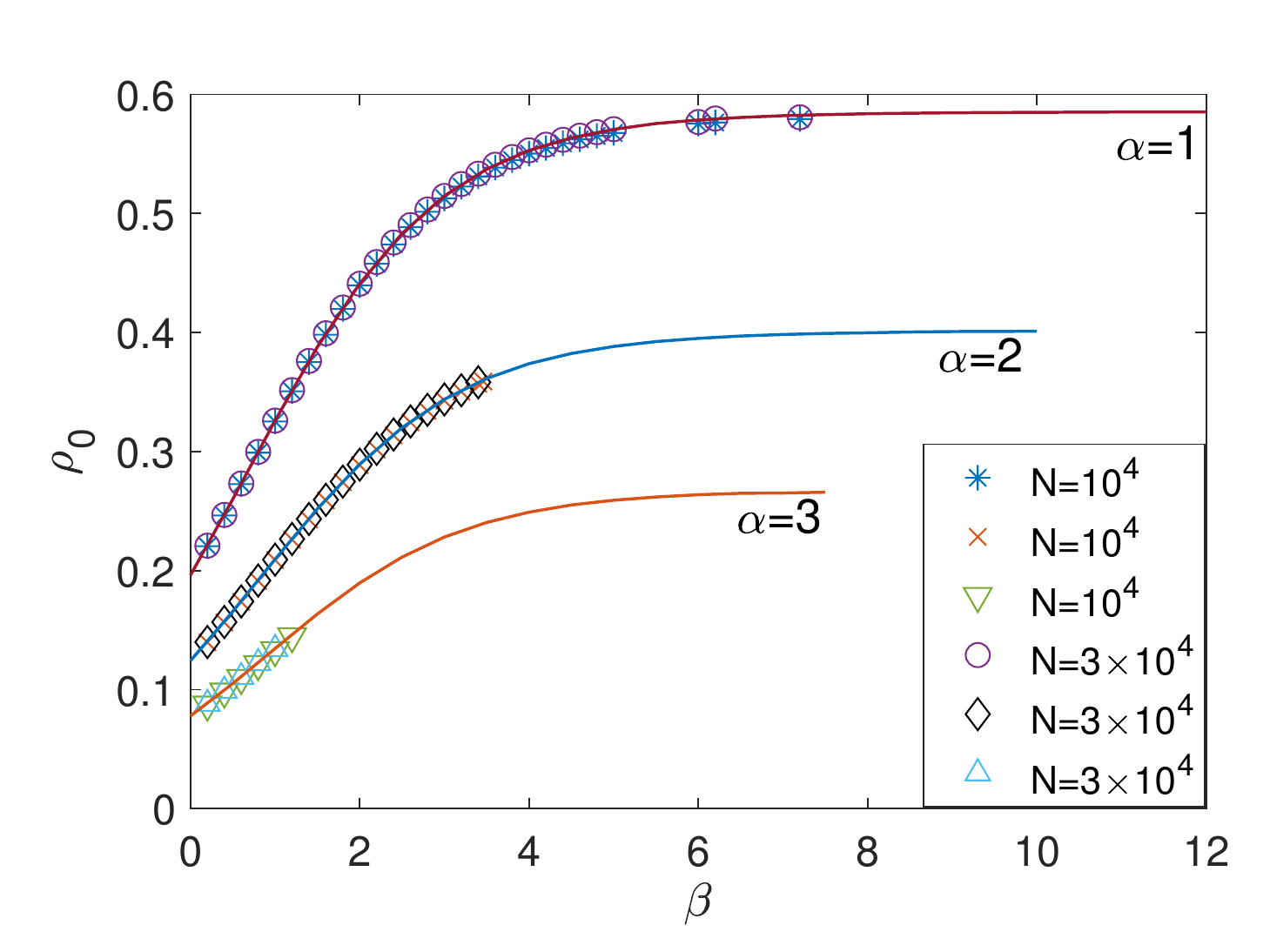}
}
\subfigure[]{
\includegraphics[width=3.0in]{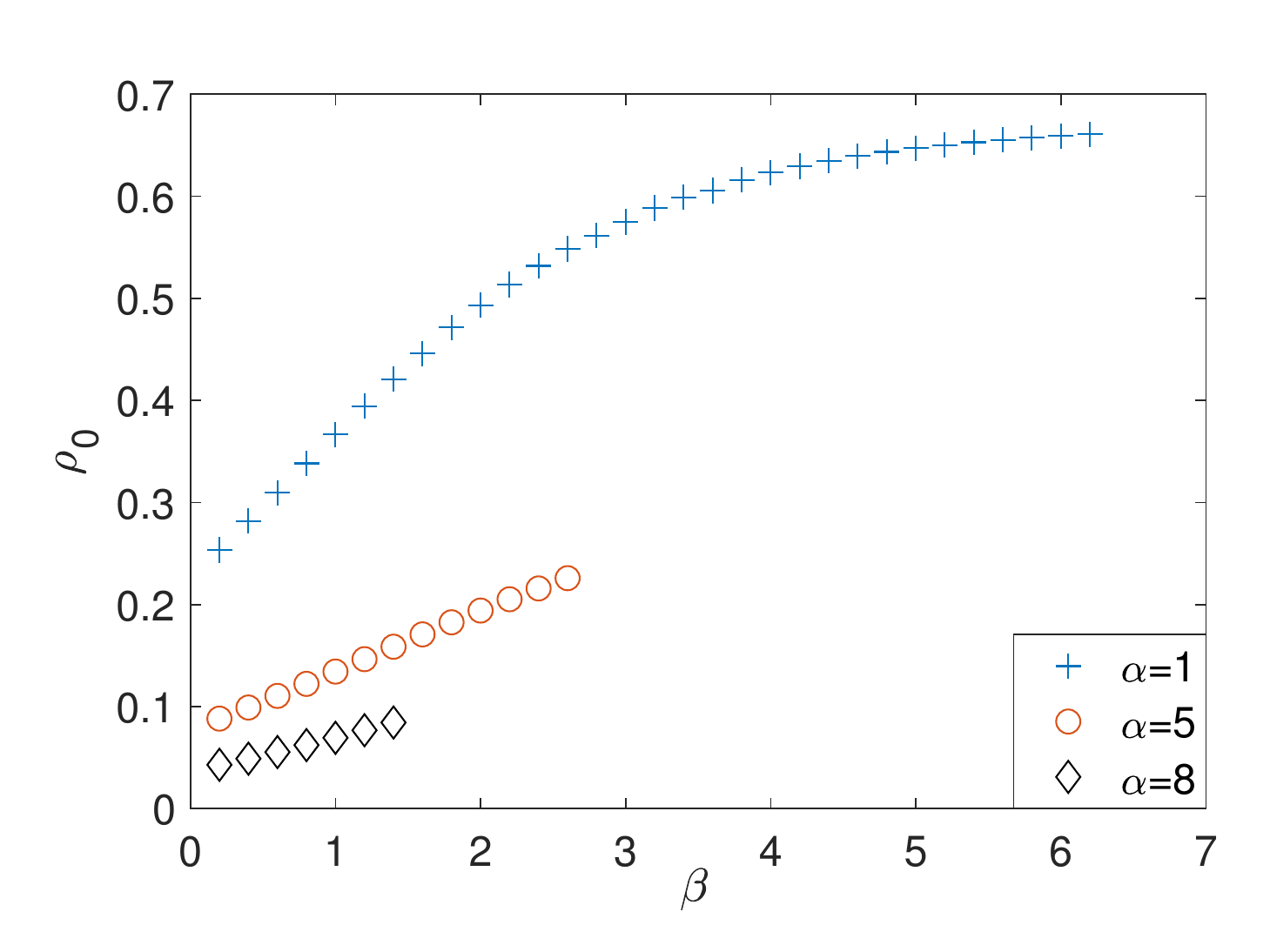}
}
\caption{
Fraction $\rho_0$ of null variables versus the inverse temperature $\beta$, as estimated by BP iteration computations on a single $K$-SAT instance of size $N$ and clause density $\alpha$ (symbols), or through RS population dynamics simulations at the ensemble level (lines). (a) $K=3$, $N = 10^4$ or $3\times 10^4$, and $\alpha = 1, 2, 3$ from top to bottom (less than the satisfiability threshold value $\alpha_s\approx 4.267$). (b) $K=4$, $N=10^4$, and $\alpha = 1, 5, 8$ from top to bottom (again less than the satisfiability threshold value $\alpha_s\approx 9.930$).  At a given value of $\alpha$ the BP iteration becomes non-convergent as $\beta$ exceeds certain critical value. At these higher $\beta$ values we can still estimate $\rho_0$ by averaging over many non-convergent BP iteration steps (data not included here).
}
\label{fig:e}
\end{figure}

We run BP iterations on single $K$-SAT instances to obtain the values of $\rho_0$ as a function of $\beta$. As initial conditions all the cavity distributions $q_{i\rightarrow a}(s_i)$ are assumed to be the uniform distribution over the three states. Some of the numerical results are shown in Fig.~\ref{fig:e}. As expected, the null fraction $\rho_0$ increases with $\beta$ for each value of clause density $\alpha$. At $\alpha = 1$ BP is always convergent (both for $K = 3$ and $K = 4$) and $\rho_0$ reaches a limiting value $\rho_0^{\rm max}$ as $\beta$ becomes large. When $\alpha$ increases, however, we find that the BP iteration is convergent only for sufficiently small $\beta$ values (e.g., up to $\beta \approx 3.46$ for the $3$-SAT instances at $\alpha=2$ and up to $\beta \approx 2.6$ for the $4$-SAT instance at $\alpha = 5$). This non-convergent behavior of BP indicates that ergodicity is broken at high $\beta$ values and the system enters into the spin glass phase as the null fraction $\rho_0$ exceeds certain threshold value~\cite{Mezard2001,Montanari2004,Montanari2008}.

\begin{figure}
\centering
\subfigure[]{
\includegraphics[width=3.0in]{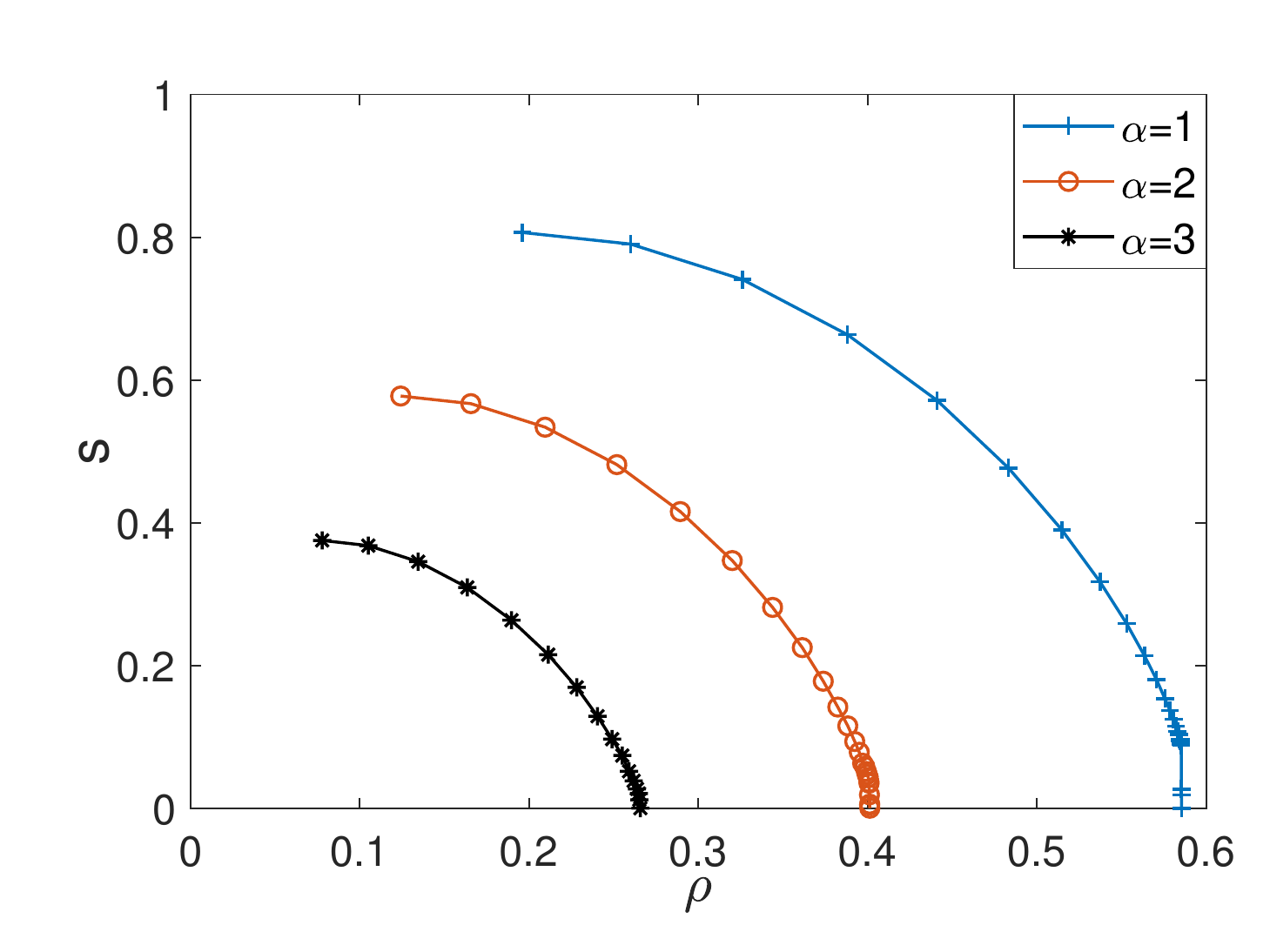}
}
\subfigure[]{
\includegraphics[width=3.0in]{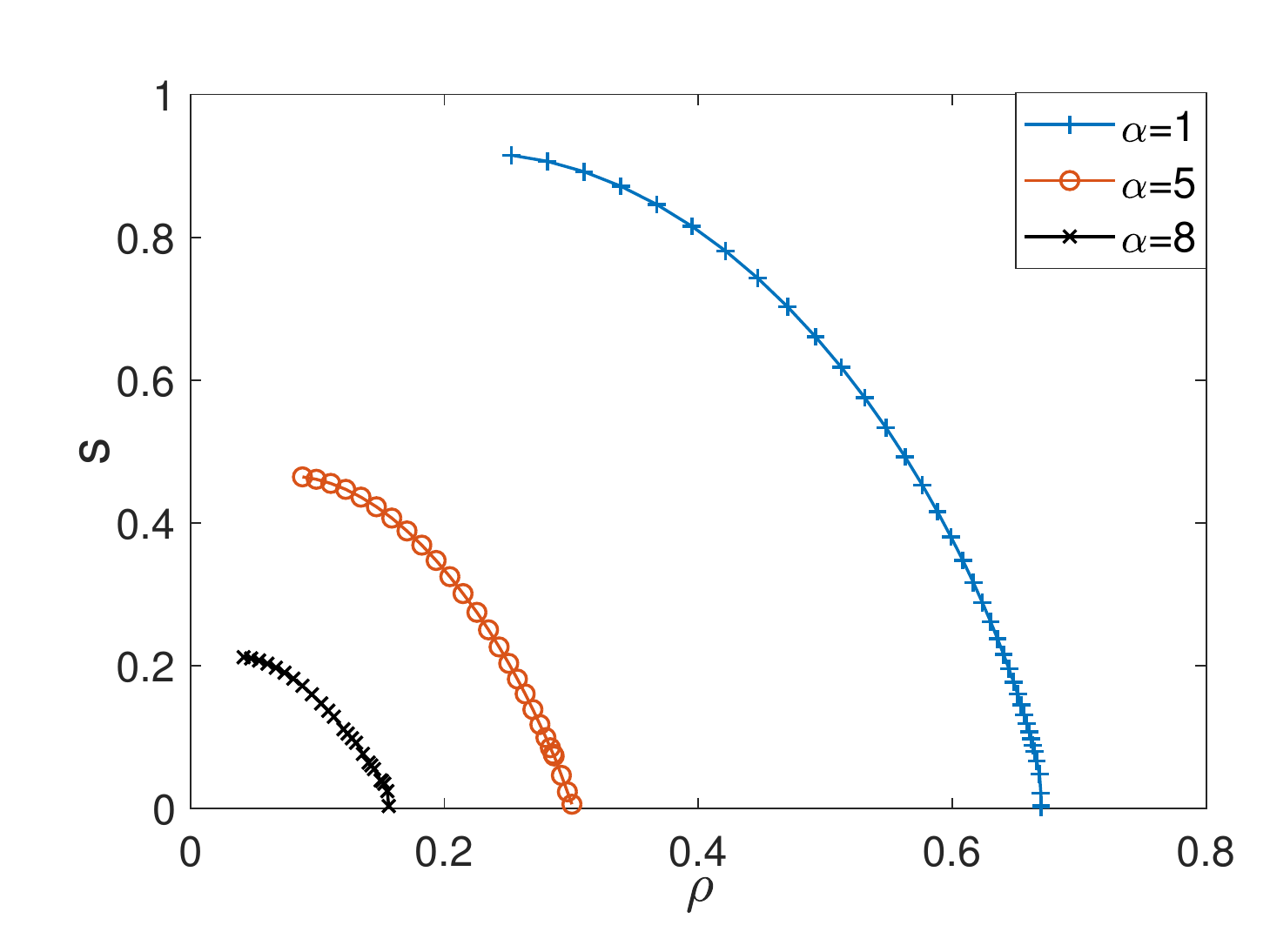}
}
\caption{
$K$-SAT entropy density $s$ versus the fraction of null variables $\rho_0$, as predicted by the RS population dynamics simulations. (a) $K=3$,  and the clause density is $\alpha = 1, 2, 3$ from top to bottom; (b) $K=4$, and $\alpha = 1, 5, 8$ from top to bottom. The maximum fraction $\rho_0^{\rm max}$ of null variables is taken to be the value of $\rho_0$ at which the entropy density $s$ reaches zero.
}
\label{fig:es}
\end{figure}

To compute the maximum null fraction $\rho_0^{\rm max}$ within the RS mean field theory, we perform population dynamics simulations on the BP equations following the standard method in the literature~\cite{Mezard2009}, and get ensemble-averaged results on $\rho_0$ and the entropy density. Again all the cavity distributions $q_{i\rightarrow a}(s_i)$ are initialized to be the uniform distribution over $s_i \in \{-1, 0, 1\}$ in this population dynamics simulation. The ensemble-averaged results of $\rho_0$ for the random $3$-SAT cases are shown in Fig.~\ref{fig:e}a, and they are in good agreement with the single-instance results. The relationship between entropy density $s$ and $\rho_0$ is shown in Fig.~\ref{fig:es}. The maximum entropy value of entropy density decreases with clause density $\alpha$. For example at $\alpha=1$ the most probable fraction of null variables is $\rho_0 \approx 0.2$ (for $K=3$), while this most probable fraction decreases to $\rho_0 \approx 0.08$ as the clause density  increases to be $\alpha = 3$.

\begin{figure}
\centering
\subfigure[]{
\includegraphics[width=3.0in]{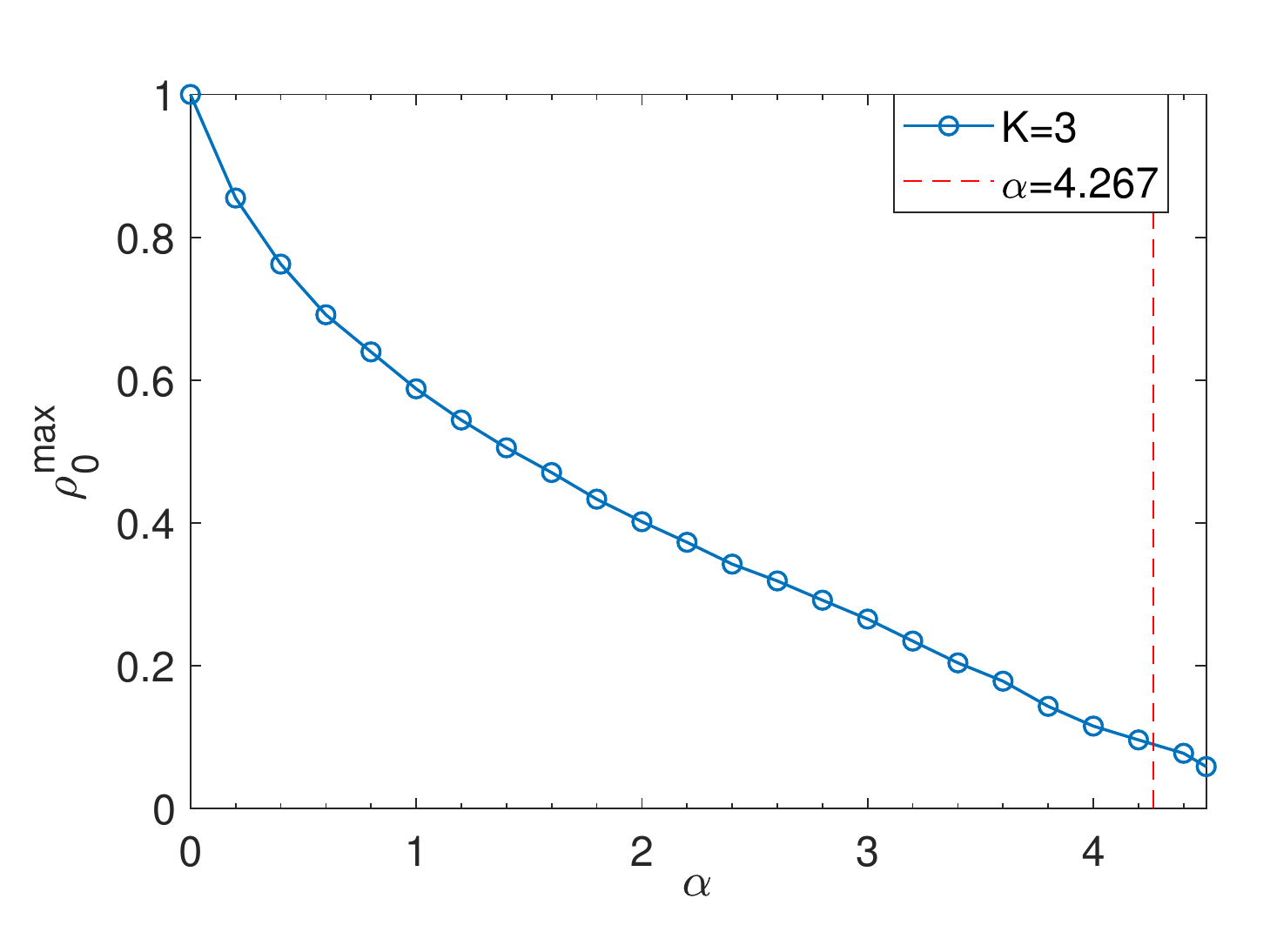}
}
\subfigure[]{
\includegraphics[width=3.0in]{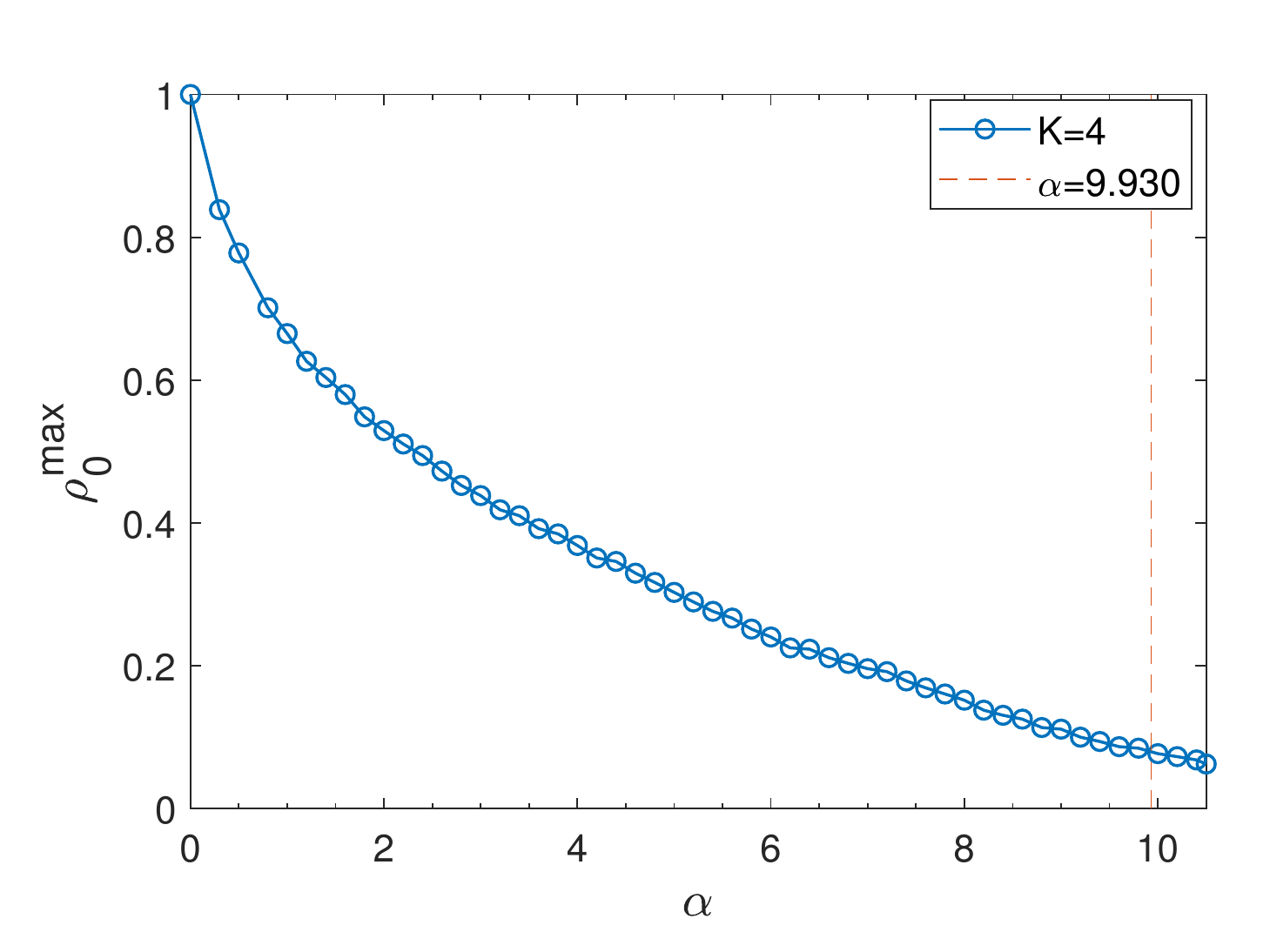}
}
\caption{
The maximum fraction $\rho_0^{\rm max}$ of null variables versus the clause density $\alpha$ of the random $K$-SAT problem, as predicted by the RS population dynamics simulations. (a) $K=3$; (b) $K=4$. The vertical dashed lines mark the predicted satisfiability phase transition point $\alpha_{s}$ ($\approx 4.267$ for $K=3$ and $\approx 9.930$ for $K=4$~\cite{Mezard2002,Mezard2002a,Mertens2006}).}
\label{fig:6}
\end{figure}

 The maximum fraction $\rho_0^{\rm max}$ of null variables also decreases with $\alpha$, as is shown in Fig.~\ref{fig:6}. This shrinking trend is fully expected. As the density of clauses increases, more and more variables need to actively participate in satisfying these clauses and so fewer variables can be spared. The true interesting feature of Fig.~\ref{fig:6} is that $\rho_0^{\rm max}$ is positive even at the satisfiability phase transition point $\alpha_{\rm s}$. This indicates that there are still an exponential number ($\sim \exp[N \rho_0^{\rm max}(\alpha_{\rm s})]$) of satisfying Ising configurations $\underline{\sigma}$ right at the transition point $\alpha_{\rm s}$. These solutions probably form a dense cluster and they are quite atypical configurations in the conventional two-state $K$-SAT model~\cite{Baldassi2015,Baldassi2016,Budzynski2019}. Our three-state model offers a simple way of emphasizing these solution clusters.
 
 In Table~\ref{tab:BPD} we compare the values of $\rho_{0}^{\rm max}$ as predicted by the RS mean field theory and the corresponding values achieved by the BPD message-passing algorithm of the next section, for the random $3$-SAT problem with clause densities $\alpha$ close to the satisfiability threshold $\alpha_{\rm s}\approx 4.267$. It's remarkable that our algorithm successfully construct solutions with null-variable fractions $\rho_0$ pretty close to the theoretical $\rho_0^{\rm max}$ values. This table also indicates the RS theoretical predictions are quite reasonable.

 \begin{table*}
\caption{
\label{tab:BPD}
Comparison of the predicted value of maximum null-variable fraction $\rho_{0}^{\rm max}$ by the RS mean field theory and the null-variable fraction $\rho_0$ of solutions constructed by the BPD algorithm, for the random $3$-SAT problem at clause densities $\alpha \in [4.0, 4.2]$. The inverse temperature is set to be $\beta \! \approx \! 5.0$. Each BPD data is the average value over ten single random $3$-SAT instances of size $N = 10^4$.}
\centering
    \begin{tabular}{l|ccccccccccc}
      \hline \hline 
       $\alpha$  & $4.00$ & $4.02$ & $4.04$ & $4.06$ & $4.08$ & $4.10$ & $4.12$ & $4.14$ & $4.16$ & $4.18$ & $4.20$ \\
       \hline 
       RS & $0.1298$ & $0.1275$ & $0.1258$ & $0.1234$ & $0.1214$ & $0.1191$ & $0.1170$ & $0.1149$ & $0.1128$ & $0.1106$ & $0.1085$ \\
       BPD & $0.1242$ & $0.1201$ & $0.1186$ & $0.1174$ & $0.1123$ & $0.1109$ & $0.1084$ & $0.1071$ & $0.1060$ & $0.1042$ & $0.0941$   \\
      \hline \hline
    \end{tabular}
\end{table*}

\section{Belief-propagation guided decimation}
\label{section:BPD}

Various local-search heuristic algorithms and message-passing algorithms have been proposed for the $K$-SAT problem (see, e.g., Refs.~\cite{Mezard2002,Montanari-etal-ARXIV-2007,Alava-etal-2008,Gomes-etal-2008,Baader2008,Braunstein2005}. Here we adopt the belief-propagation guided decimation (BPD) algorithm, widely used in the $K$-SAT problem and other optimization problems such as minimum feedback vertex set and minimum vertex covers~\cite{Mugisha2016,Montanari-etal-ARXIV-2007,Zhou-2013,Zhao-Zhou-2014},  to construct satisfying configurations with close-to-maximum number of null variables.

The marginal state probabilities $q_i(s_i)$ for all the variables of a given problem instance are estimated by BP iterations at  certain fixed value of inverse temperature $\beta$, and the algorithm then fix a small fraction $r$ of the variables $i$ to $+1$ and $-1$ spin values based on these marginal probability distributions. The marginal state probabilities of the remaining variables are then updated by BP again and then a small fraction $r$ of the remaining variables are fixed to $+1$ or $-1$ spin values. The algorithm stops as soon as a satisfying configuration has been reached or a conflict has been encountered (i.e., the formula can not be satisfied by the remaining variables). In the former case the variables not yet assigned any spin value form the set of null variables.

In our implementation, the variables $i$ to be fixed in one round of the BPD decimation process are those whose marginal probability values $q_i(+1)$ or $q_i(-1)$ are the largest among all the variables not yet fixed. For such a variable $i$, we set its spin to $\sigma_i = +1$ if $q_i(+1) > q_i(-1)$, otherwise its spin is set to $\sigma_i = -1$. After variable $i$ is fixed we then simplify the formula by deleting the clauses that are satisfied by variable $i$ and fix some additional variables $j$ if they are required to be fixed by some other clauses.

\begin{figure}
\centering
\includegraphics[width=3.0in]{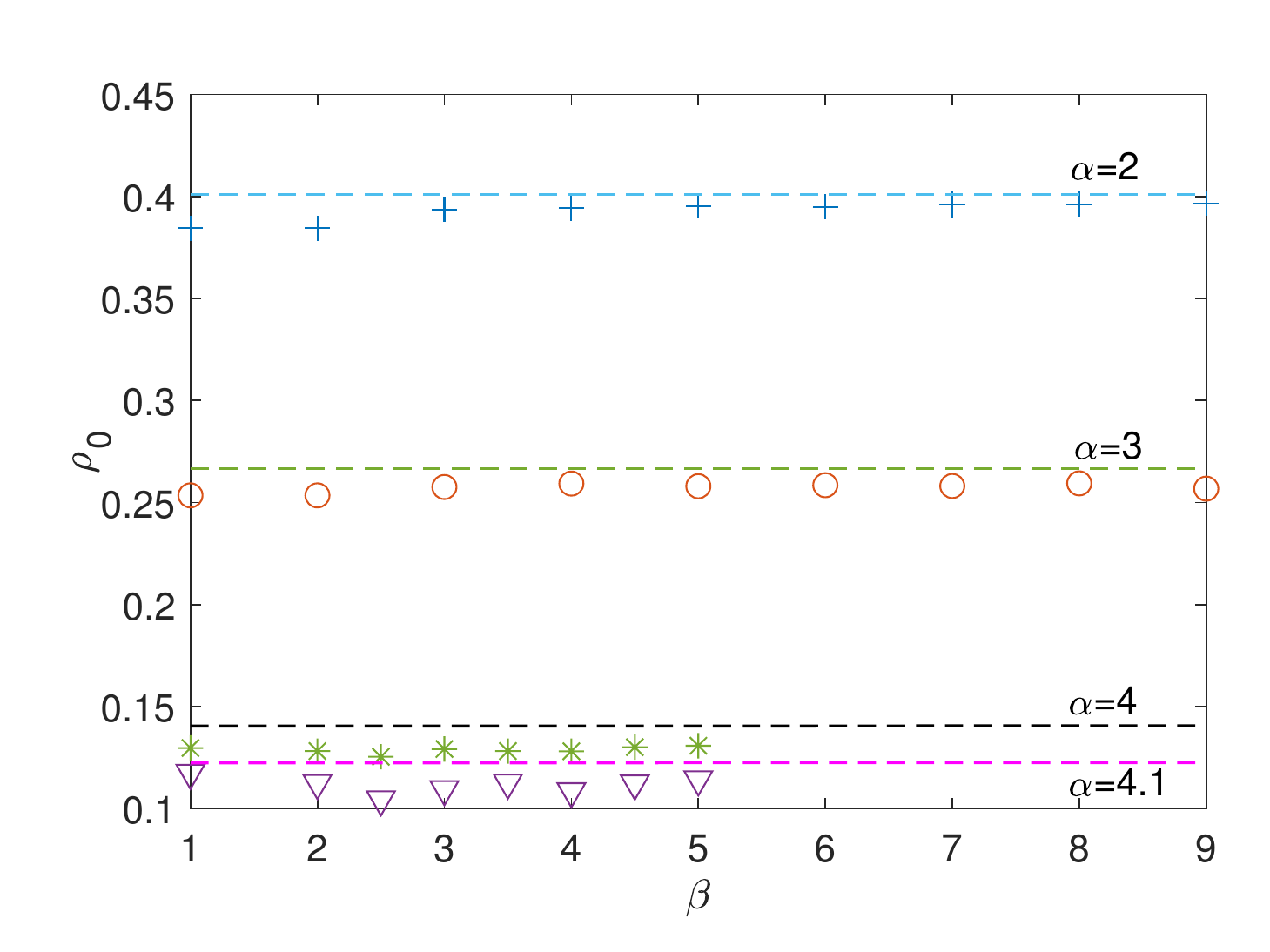}
\caption{
The mean fraction $\rho_0$ of null variables in satisfying configurations of random $3$-SAT instances, obtained by the BPD algorithm with fixed inverse temperature $\beta$. A single problem instance of size $N=10^4$ is used and its clause density is $\alpha = 2, 3, 4, 4.1$ from top to bottom. The horizontal dashed lines are the corresponding maximum fraction $\rho_0^{\rm max}$ of null variables as predicted by the RS mean field theory. 
}
\label{fig:BPDbeta}
\end{figure}

Some of the results obtained by BPD for single random $3$-SAT instances are shown in Fig.~\ref{fig:BPDbeta}. At each value of clause density $\alpha$ we find that the density $\rho_0$ of null variables obtained by BPD is quite close to the theoretically predicted maximum value $\rho_0^{\rm max}$. This fact is also demonstrated in Table~\ref{tab:BPD}. Another interesting feature is that the BPD performance is only slightly dependent on the value of the inverse temperature $\beta$. 

\begin{figure}
\centering
\subfigure[]{
\includegraphics[width=3.0in]{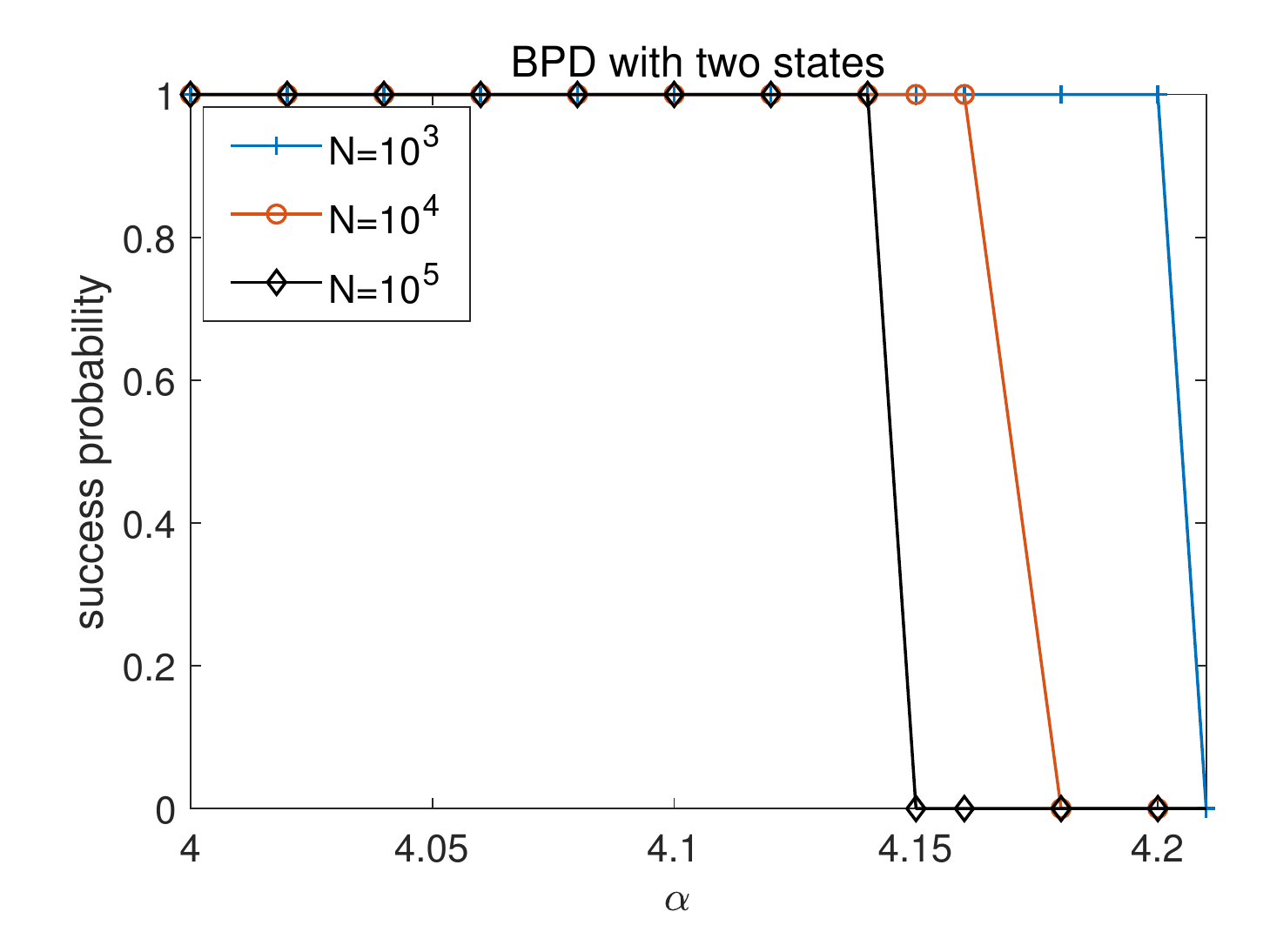}
}
\subfigure[]{
\includegraphics[width=3.0in]{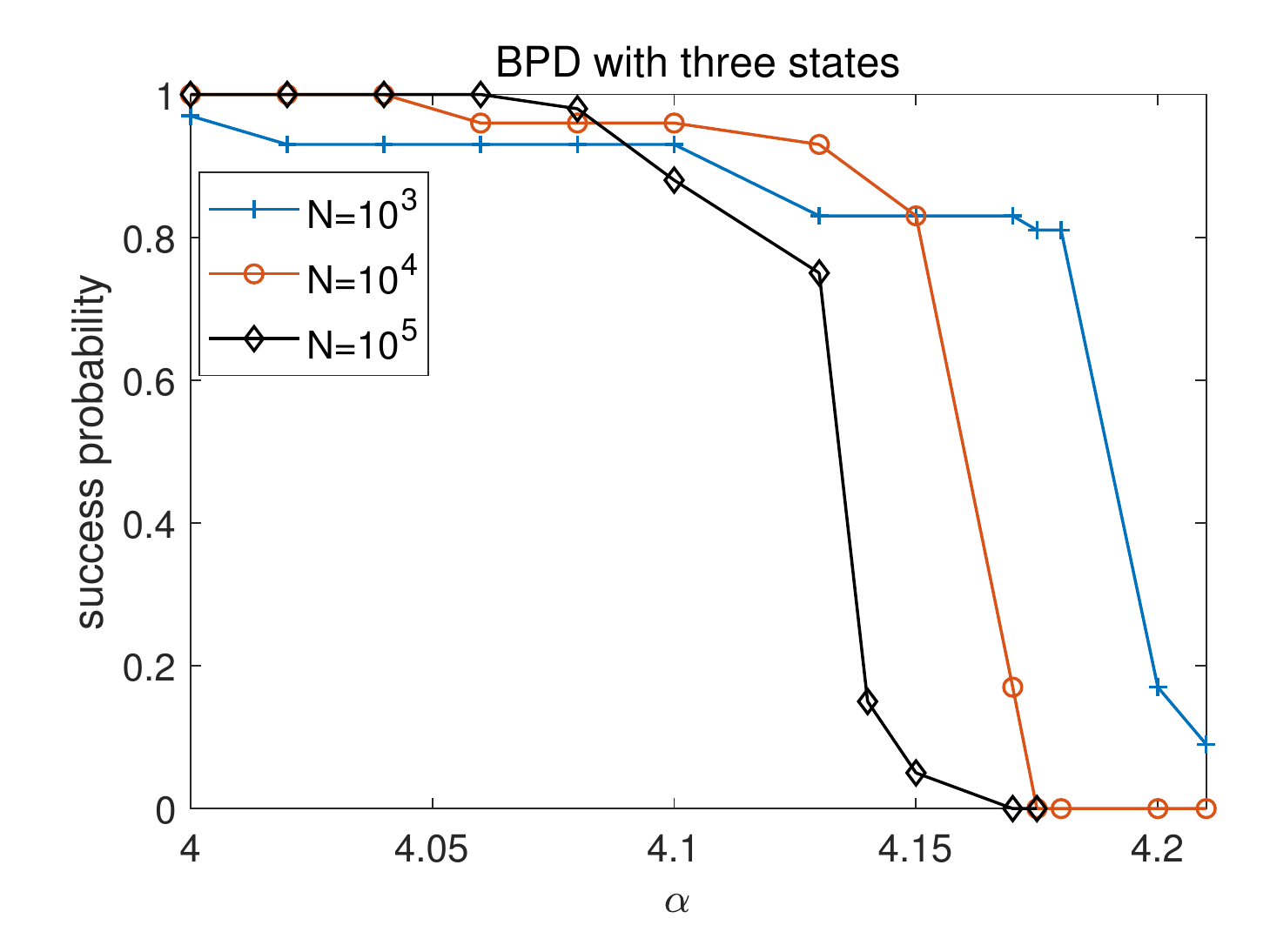}
}
\caption{
Comparison on the performance of the BPD algorithm with two states (a) or three states (b). Three random $3$-SAT formulas of size $N \in \{10^3, 10^4, 10^5\}$ are used in this comparison, and the first $\alpha N$ clauses of these instances are considered. Each data point shows the probability of successfully constructing a solution by the BPD algorithm among $1000$ independent trials. The BPD algorithms are not sensitive to the inverse temperature $\beta$ (see Fig.~\ref{fig:BPDbeta}), so we simply set it to a moderate value, e.g., $\beta = 5.0$.
}
\label{fig:8}
\end{figure}

When the clause density $\alpha$ is approaching the satisfiability threshold $\alpha_{\rm s}$, it become more and more difficult for the BPD algorithm to construct satisfying configurations. This happens both for the conventional BPD algorithm with only two states~\cite{Montanari-etal-ARXIV-2007} and the present one with three states. We notice that the failure of conventional two-state BPD occurs more abruptly than the present three-state BPD algorithm (Fig.~\ref{fig:8}). For example, when applying the two-state BPD algorithm on a $3$-SAT instance of size $N=10^5$ and clause density $\alpha = 4.15$, the two-state BPD always fails in $1000$ independent repeats, while our three-state BPD algorithm succeeds to constructing satisfying configurations in about $8$ out of $1000$ independent trials. The gradual declining behavior shown in Fig.~\ref{fig:8}b for the three-state BPD algorithm may indicate its applicability for the most hard problem instances.

It may be possible to further improve the performance of BPD by modifying the spin fixation process. We could check the effect of fixing the state of a variable $i$. If this fixation does not force any additional spin fixation we accept it; otherwise we may accept it with some small probability, favoring fixation choices which have the least effects to the remaining variables.

\section{Belief-propagation guided reinforcement}
\label{section:BPRf}

We also implement a slightly different message-passing algorithm, belief-propagation guided reinforcement (BPR), for constructing maximally flexible satisfying configurations.  The main idea of BPR is the same as that of BPD but it allows each variable state $s_i$  to be modified multiple times during the search process, so it is much more robust~\cite{Braunstein-Zecchina-2006}. The most important feature of BPR is the memory effect which is realized by an external reinforcement vector $\underline{\psi}_i \equiv (\psi_i^{-1}, \psi_i^0, \psi_i^{+1})$ on every variable $i$. With this reinforcement factor $\underline{\psi}_i$, the BP equation (\ref{a}) is slightly modified as
\begin{equation}
\label{BP3}
q_{i\to a}(s_i) = \frac{\psi_i^{s_i} e^{\beta\delta_{s_i}^0} \prod\limits_{{b\in\partial i\backslash a}} p_{b\to i}(s_i)}
{\sum\limits_{s_i^\prime}\psi_i^{s_i^\prime} e^{\beta\delta_{s_i^\prime}^0}\prod\limits_{b\in\partial i\backslash a}p_{b\to i}(s_i^\prime)} \; .
\end{equation}
The expression (\ref{q}) for the marginal probability $q_i(s_i)$ is also revised to be
\begin{equation}
q_{i}(s_i) = \frac{\psi_i^{s_i} e^{\beta \delta_{s_i}^0} \prod\limits_{{a\in\partial i}} p_{a\to i}(s_i)}
{\sum\limits_{s_i^\prime}\psi_i^{s_i^\prime}e^{\beta \delta_{s_i^\prime}^0} \prod\limits_{a\in\partial i} p_{a\to i}(s_i^\prime) } \; .
\label{BPRequation}
\end{equation}
\begin{figure}
\centering
\includegraphics[width=3.0in]{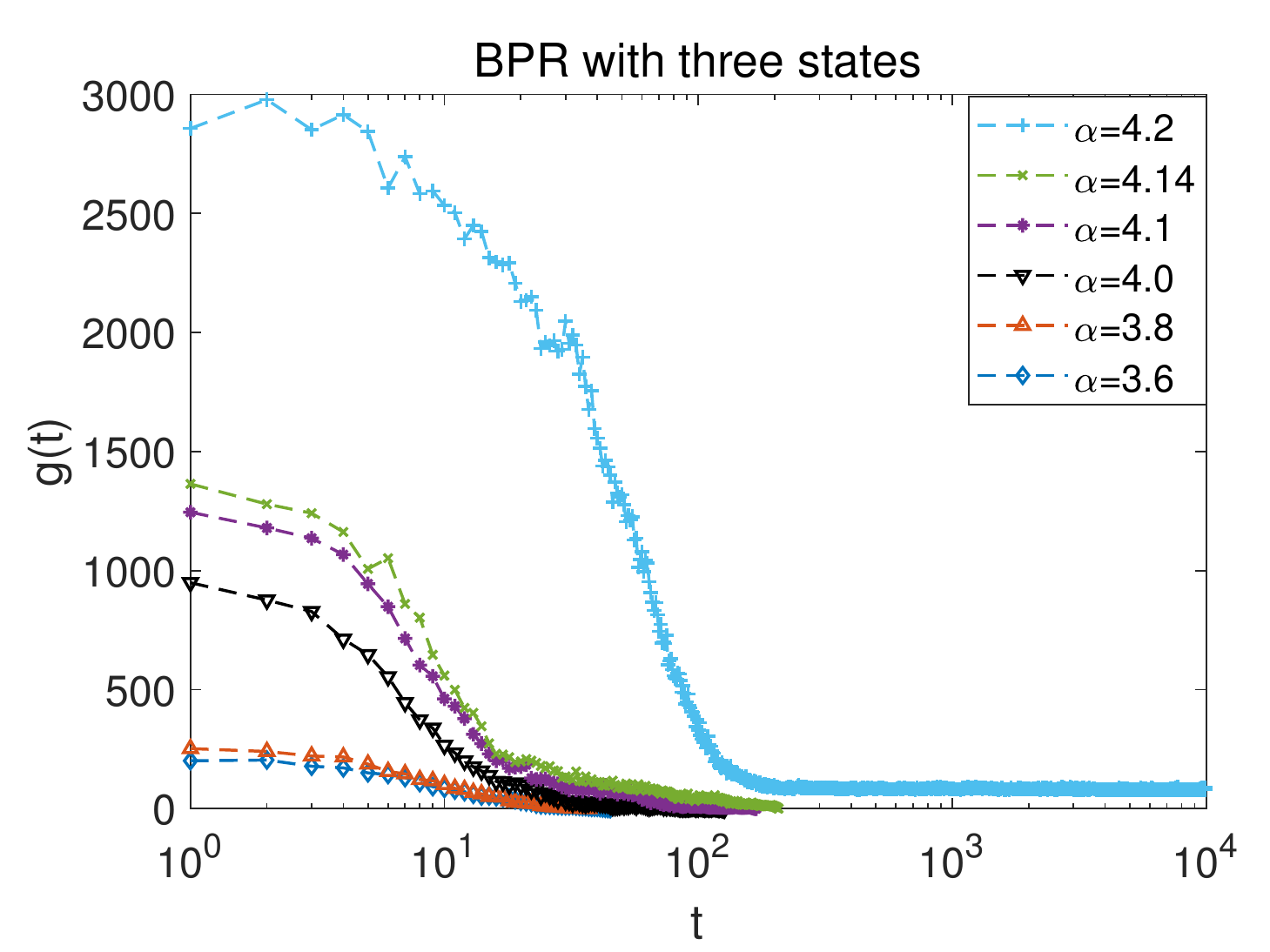}
\caption{
The performance of the BPR algorithm on single random $3$-SAT instances of size $N = 10^4$ and different fixed clause densities $\alpha$ ranging from $3.6$ to $4.2$. Each data set (symbols linked by lines) shows the evolution of the number $g(t)$ of unsatisfied clauses with the evolution time $t$ of BPR. The inverse temperature is $\beta \! \approx \! 5.0$.
}
\label{fig:BPR}
\end{figure}

We implement BPR as follows. First, the memory vector of each variable $i$ is initialized as $\underline{\psi}_i = (1, 1, 1)$, i.e., $\psi_i^{s_i} =1$ for $s_i\in \{-1, 0, 1\}$. Then at each BPR time step $t$: (1) we first iterate equations (\ref{BP3}) and (\ref{b}) on all the links of the factor graph a number of repeats (e.g., ten times) and; (2) then compute the marginal probability distributions $q_i(s_i)$ following Eq.~(\ref{BPRequation}); and (3) then update the memory vectors $\underline{\psi}_i$ of all the variables according to $q_i(s_i)$. We take the following reinforcement rule: If the most likely state of variable $i$ is $s_i^* \in \{-1, 0, +1\}$, that is, if $p_i(s_i^*)$ is larger than the other two probabilities, then a small amount $\eta$ is added to $\psi_i(s_i^*)$ while the other two elements of $\underline{\psi}_i$ are not changed (we set $\eta=0.002$).

At each BPR time step $t$ we also assign an instantaneous spin value $\sigma_i(t) \in \{ -1, +1\}$ to each variable $i$. If the most likely state of $i$ is predicted to be $+1$ (or $-1$) by the marginal probability distribution $q_i(s_i)$, then we set $\sigma_i(t) = +1$ (and respectively, $\sigma_i(t)=-1$); if the null state $s_i=0$ is the most likely state of $i$, we set $\sigma_i(t) = 1$ (respectively, $\sigma_i(t)=-1$) if this variable has more links of positive (respectively, negative) coupling constants. After an instantaneous spin configuration is obtained by this way, we then count the total number of unsatisfied clauses and denote this number as $g(t)$.

Some evolution trajectories of $g(t)$ with BPR time $t$ are shown in Fig.~\ref{fig:BPR} for a single random $3$-SAT formula. When the clause density $\alpha$ is less than $4.15$ we find that $g(t)$ reaches zero in less than $10^3$ steps. When $\alpha = 4.2$ (quite close to $\alpha_{\rm s}$), however, we find that $g(t)$ saturates to a small positive value, meaning the algorithm fails to reach a satisfying solution.

\section{CONCLUSION}
\label{section:concl}

In summary, we designed a three-state spin glass model to explore the atypical and maximally flexible solutions of the random $K$-SAT problem. We succeeded to construct satisfying solutions with a close-to-maximum fraction $\rho_0$ of null variables for single random $K$-SAT instances, when the clause density $\alpha$ of these instances are not too close to the satisfiability phase transition point $\alpha_{\rm s}$. Our replica-symmetric mean field theoretical results suggested that even at $\alpha = \alpha_{\rm s}$ the maximum fraction $\rho_0^{\rm max}$ of null variables is still positive.

The RS mean field theory adopted in this work is clearly inadequate when the inverse temperature $\beta$ becomes large and the clause density $\alpha$ is approaching $\alpha_{\rm s}$. The non-convergence of the BP iterations might have contributed to the failure of the BPD and BPR algorithms at the hard region of $\alpha \approx \alpha_{\rm s}$. We need to extend the mean field theory to the 1RSB level and get refined predictions on the maximum null-variable fraction $\rho_0^{\rm max}$ and improved message-passing algorithms. It would be very interesting to see whether the 1RSB theory still predicts a positive $\rho_0^{\rm max}$ at $\alpha_{\rm s}$.  We will address these issues in a follow-up report.

\begin{acknowledgments}
One of authors(Han Zhao) thanks Yi-Zhi Xu for helpful discussions. This work was supported by the National Natural Science Foundation of China Grants No.11975295 and No.11947302, and the Chinese Academy of Sciences Grant No.QYZDJ-SSW-SYS018. Numerical simulations were carried out at the HPC cluster of ITP-CAS.
\end{acknowledgments}

\end{document}